\documentstyle[12pt]{article}
  \hoffset = - 1 truecm
\def\beq{\begin{equation}}  \def\eeq{\end{equation}}
\def\bea{\begin{eqnarray}}  \def\eea{\end{eqnarray}} \def\nn{\nonumber}
\def\noi{\noindent} \def\beeq{\begin{eqnarray}}
\def\eeeq{\end{eqnarray}}
\def\lsim{\raise0.3ex\hbox{$<$\kern-0.75em\raise-1.1ex\hbox{$\sim$}}}
\def\gsim{\raise0.3ex\hbox{$>$\kern-0.75em\raise-1.1ex\hbox{$\sim$}}}

\newcommand\mysection{\setcounter{equation}{0}\section}
\renewcommand{\theequation}{\thesection.\arabic{equation}}
\newcounter{hran} \renewcommand{\thehran}{\thesection.\arabic{hran}}

\def\bmini{\setcounter{hran}{\value{equation}}
\refstepcounter{hran}\setcounter{equation}{0}
\renewcommand{\theequation}{\thehran\alph{equation}}\begin{eqnarray}}

\def\bminiG#1{\setcounter{hran}{\value{equation}}
\refstepcounter{hran}\setcounter{equation}{-1}
\renewcommand{\theequation}{\thehran\alph{equation}}
\refstepcounter{equation}\label{#1}\begin{eqnarray}}

\def\emini{\end{eqnarray}\relax\setcounter{equation}{\value{hran}}\renewcommand{\theequation}{\thesection.\arabic{equation}}}

\def\ben{\begin{enumerate}}  \def\een{\end{enumerate}}

\def\cite#1{[\ref{#1}]} 
\def\citd#1#2{[\ref{#1}, \ref{#2}]}

\def\citm#1#2{[\ref{#1}--\ref{#2}]}

\def\citl#1#2#3{[\ref{#1}--\ref{#2}, \ref{#3}]}

\begin{document} 

\centerline{\Large\bf Generalized Gravitational S-Duality and
} \par \vskip 3 truemm 
\centerline{\Large\bf the Cosmological Constant Problem} \vskip 1 truecm

\begin{center} {\bf Ulrich Ellwanger}\footnote{E-mail :
Ulrich.Ellwanger@th.u-psud.fr}\par \vskip 5 truemm

Laboratoire de Physique Th\'eorique\footnote{Unit\'e Mixte de Recherche
- CNRS - UMR 8627}\par Universit\'e de Paris XI, B\^atiment 210,
F-91405 Orsay Cedex, France\par \vskip 5 truemm

\end{center} \vskip 2 truecm

\begin{abstract} We study $S$-duality transformations that mix the
Riemann tensor with the field strength of a 3-form field.  The dual of
an $(A)dS$ space time -- with arbitrary curvature -- is seen to be flat
Minkowski space time, if the 3-form field has vanishing field strength
before the duality transformation. It is discussed whether matter could
couple to the dual metric, related to the Riemann tensor after a
duality transformation. This possibility is supported by the facts that
the Schwarzschild metric can be obtained as a suitable contraction of
the dual of a Taub-NUT-AdS metric, and that metrics describing FRW
cosmologies can be obtained as duals of theories with matter in the
form of torsion. \end{abstract}

\vskip 3 truecm \noi LPT Orsay 04-101 \par \noi October 2004 \par

\newpage 
\pagestyle{plain} 
\baselineskip 20pt

\mysection{Introduction} \hspace*{\parindent} 
The cosmological constant problem has motivated various attempts to
modify Einstein's theory of gravity. In the present paper we study
$S$-duality acting on the gravitational field. The presence of a dual
graviton is motivated by hidden symmetries in $d=11$
supergravity/M-theory \cite{west}. \par

Subsequently we confine ourselves to $d=4$ dimensions, where
the dual of a graviton is again a graviton-like symmetric two-component
tensor field \citm{3r}{henn}. We want to persue the question whether a
metric obtained through such a duality transformation can describe
naturally a space time which is flat (rather than (A)dS), although the
original space time (before the duality transformation) is (A)dS with
arbitrary cosmological constant. The idea is that, although space-time
is possibly strongly (A)dS in one version of gravity, we ``see'' its
dual that is obtained by the above duality transformation.\par

In order to describe the concept behind this approach it is best to
start with the well understood concept of duality in electromagnetism.
If $A$ is an abelian (one form) gauge field, $F$ its (two form) field
strength, the duality transformation relating $A$ and $F$ to its duals
$\widetilde{A}$ and $\widetilde{F}$ reads

\beq\label{1.1e}  
\widetilde{F}(\widetilde{A}) = \star F(A)\ .
\eeq

Whereas a tensor $\widetilde{F}$ can always be defined through eq.
(\ref{1.1e}), it can be expressed in terms of a dual gauge field 
$\widetilde{A}$ only if $d\widetilde{F}=0$, i.e. if $F(A)$ satisfies
the (free) equations of motion $d\star F = 0$. 

The introduction of the dual gauge fields $\widetilde{A}$ would be
particularly useful, if magnetic monopoles would exist: It is not
possible to couple a magnetic monopole (described by a magnetic current
$J_M$) locally to $A$. On the other hand, such a magnetic current
couples locally to $\widetilde{A}$ like $\widetilde{A}\cdot J_M$ in the
same way as an electric current couples to $A$. If only magnetic
monopoles would exist, but no electrically charged particles, it would
be reasonable to replace $A$ by $\widetilde{A}$ everywhere in the
theory. Since electromagnetism is self dual, the resulting theory
(written in terms of $\widetilde{A}$) would look the same as a theory
with electrically charged particles written in terms of $A$, and one
would simply denote the original magnetic monopoles as electric
monopoles.

Already in the relatively simple case of electromagnetism it is not
quite trivial to find a parent action $S(A,\widetilde{A})$ that allows
to obtain the duality relation (\ref{1.1e}) as a consequence of the
equations of motion: Such a parent action is either not manifestly
Lorentz covariant \citd{Zwanziger:1970hk}{Schwarz:1993vs} or includes 
additional auxiliary fields \citd{Witten:1995gf}{Pasti}. 

On the other hand, a parent action is not necessary in order to define
a duality transformation relating $A$ and $\widetilde{A}$: It is
sufficient to define such a relation via (\ref{1.1e}), and to verify
that equations of motion and Bianchi identities for $F$ and
$\widetilde{F}$ get interchanged.

In the present paper we persue the second approach in order to define
dual gravity. Subsequently it will be useful to work with tensors with
(latin) indices in (flat) tangent space that are related to tensors
with (greek) space time indices, as usual, by contractions with a
vierbein. A duality relation analogous to (\ref{1.1e}) reads then
\citl{3r}{5r}{henn}

\beq\label{1.2e}  
\widetilde{R}_{abcd}(\widetilde{e}) = {1 \over 2} \  \varepsilon_{abef}
\ R^{ef}_{\phantom{ab}cd}(e) + \dots
\eeq

\noi where the Riemann tensor $R_{abcd}$ depends on a vierbein $e$ (or
a metric $g$) as usual, and the dual Riemann tensor
$\widetilde{R}_{abcd}$ is assumed to depend similarly on a dual
vierbein $\widetilde{e}$ (or a dual metric $\widetilde{g}$) (the dots in
(\ref{1.2e}) are introduced for later porposes).

The standard coupling of gravity to matter is of the form

\beq\label{1.3e}
G_{ab} = - T_{ab}
\eeq

\noi where $G_{ab}$ is the Einstein tensor, and $T_{ab}$ the stress
energy tensor of matter. For most practical purposes (tests of general
relativity) it suffices to consider matter in the form of point like
particles. Either such matter serves as a ``source" for the
gravitational field (as the Schwarzschild solution), or as test
particles: From the covariant conservation of the stress energy tensor
$T_{ab}$ (which has its origin in the Bianchi identity for the Riemann
tensor $R_{abcd}$) it follows that any point like object, that couples
like (\ref{1.3e}) to gravity, moves along geodesics corresponding to
the metric $g$.

Also stress energy tensors of the Friedmann-Robertson-Walker form, that
involve a matter density $\rho(t)$ and a pressure $p(t)$, can be
understood as suitable averages over pointlike sources \cite{kras}.

The cosmological constant problem arises from contributions of the form
$\sim\ \eta_{ab}\Lambda$ to $T_{ab}$, which have its origin in the
minimal coupling of fields (the fields of the standard model) to
gravity: only then potentials of classical scalar fields (Higgs
fields), vacuum condensates (as in QCD) and quantum contributions to
the expectation value of $T_{ab}$ generate values of $\Lambda$ that
gives rise, via (\ref{1.3e}), to an unobserved space time curvature 
$\sim \Lambda$. We are very far, however, from being able to test the
minimal coupling of the fields of the standard model to gravity.

Now, if dual gravity in the sense of eq. (\ref{1.2e}) exists, it would
be interesting to investigate whether macroscopic matter, that is
involved in classical tests of general relativity, may couple to a dual
metric $\widetilde{g}$ in the form

\beq\label{1.4e}
\widetilde{G}_{ab} = - \widetilde{T}_{ab}\ .
\eeq

\noi Then, since $\widetilde{T}_{ab}$ is covariantly conserved with the
dual metric in the covariant derivative, matter would move on geodesics
corresponding to the dual metric $\widetilde{g}$, and we would ``see" a
space time metric $\widetilde{g}$.

Clearly, here we have to assume that the duality transformation
(\ref{1.2e}) is accompagnied by a duality transformation acting on the
stress energy tensor such that the dual stress energy tensor is
consistent with (\ref{1.4e}). We can not expect, on the other hand,
that duality transformations including gravity -- beyond the linearized
level -- can be derived from a local parent action that would allow to
derive the contributions of matter fields to the dual stress energy
tensor from a variational principle.  We will thus proceed with the
assumption that a more complete (probably non local) consistent duality
transformation -- leading to a stress energy tensor consistent with
(\ref{1.4e}) -- exists and use, for the time being, the results of the
duality transformations (\ref{1.2e}) to define the action of duality on
the stress energy tensor. (We return to this issue in sections 3 and,
in the particular case $\widetilde{T}_{ab}=0$, in section 6.)

In the following we want to investigate, whether eq. (\ref{1.4e}) 
contradicts eq. (\ref{1.2e}) (leaving aside a possible microscopic
origin of eq. (\ref{1.4e})): At first sight it seems that we just have
to rename  $\widetilde{g}$ to $g$ in order to recover standard Einstein
gravity from eq. (\ref{1.4e}). However, if we require simultaneously the
validity of eq. (\ref{1.2e}), it is not clear whether this does not
restrict the possible configurations of $\widetilde{g}$ to an
inacceptable level.

It is already known that, in the vacuum (where $T_{ab}=
\widetilde{T}_{ab} = 0$) and in the weak field limit, this is not the
case \citd{3r}{4r}: in this limit gravity is self dual, i.e. for every
Riemann tensor on the right hand side of eq. (\ref{1.2e}) that solves
the vacuum equations of motion (whose Ricci tensor vanishes), one can
define a dual Riemann tensor, derivable from a dual metric, that solves
the vacuum equations of motion as well. (For a corresponding parent
actions quadratic in the fields see, e.g., refs. \citm{5r}{henn} and,
in the case of the Macdowell-Mansouri formalism, ref. \cite{3r}. In the
latter case, where a cosmological constant is present, it is argued
that the duality transformation leads to an inversion of the
cosmological constant.)

As in the case of Yang-Mills theories the real problems arise, however,
at the nonlinear level: At the nonlinear level we do not know how to
map $g \to \widetilde{g}$ such that the corresponding Riemann tensors
satisfy eq. (\ref{1.2e}), and such that equations of motion and Bianchi
identities get interchanged. On the other hand, only at the nonlinear
level the covariant derivatives (with respect to which the Bianchi
identities hold, and hence with respect to which the corresponding
Einstein tensors and hence the corresponding stress energy tensors are
conserved) involve the connections of the  corresponding metrics $g$ or
$\widetilde{g}$, and imply thus the motion of matter on corresponding
geodesics.

Attempts to proceed via parent actions beyond the linearized level run
into conflicts with no-go theorems on interacting theories (with at
most two derivatives) of ``dual'' gravitons \cite{beka}. Therefore we
will confine ourselves to duality relations at the level of equations
of motion (and Bianchi identities) of the form of eq. (\ref{1.2e})
allowing eventually for additional contributions on the right hand
side. We will investigate whether particular configurations of a dual
metric $\widetilde{g}$, that are of confirmed phenomenological
relevance (the Schwarzschild metric, and Freedman-Robertson-Walker
(FRW) like cosmologies), are consistent with a suitably generalized
(see below) duality transformation at the nonlinear level.

In general, once several gauge fields (and hence corresponding field
strengths) are present in a theory, these can mix under duality
transformations \citd{crem}{west} if their rank is appropriate.
Examples are gauge fields of higher rank in d=10 and d=11 supergravity
theories \citd{crem}{west}. In \cite{west} it has been proposed that
the corresponding algebras include gravity in a nontrivial way. It
seems then possible that also the field strength of gravity (the
Riemann tensor) mixes with other field strengths under a duality
transformation. Indeed, a three form gauge field (with a four form
field strength) is present in these higher dimensional supergravity
theories \citd{crem}{west}, which has the appropriate rank.

Below we show that a modification of the duality transformation
(\ref{1.2e}), that mixes the Riemann tensor with a four form field
strength, is consistent at the same level (linear in the gravitational
excitations) as duality in pure gravity, in the sense that it exchanges
equations of motion and Bianchi identities.

Moreover, this modified duality transformation is seen to have several
virtues: Notably, it transforms the Riemann tensor of a space time
metric $g$ with arbitrary curvature ((A)dS space times) into a Riemann
tensor of a space time $\widetilde{g}$ whose curvature is given by the
vev of the four form field strength before the duality transformation.
If this vev vanishes, the dual space time described by $\widetilde{g}$
is thus flat (Minkowski), irrespectively of the curvature of the
original space time described by $g$. This may point towards an
unconvential solution of the cosmological constant problem, if
macroscopic matter couples to dual gravity as discussed above. 

This mechanism to obtain a vanishing cosmological constant is very
different from its cancellation by a specific (fine tuned) value for
the vev of the four form field strength  as considered in \cite{6r},
and also from the proposal in \cite{7r}: Here we suggest that, although
space-time is possibly strongly (A)dS in one version of gravity, we
``see'' its dual that is obtained by the above duality
transformation.\par

Next, it is impossible to obtain a Schwarzschild metric $\widetilde{g}$
as the dual of a metric $g$ (that solves the vacuum equations of
motion) using the ``standard" gravitational duality transformation. At
the linearized level, a Schwarzschild metric $\widetilde{g}$ with mass 
$\widetilde{m}$ is dual to a Taub-NUT metric $g$ with NUT parameter
$\ell =  \widetilde{m}$ (and mass $m = 0$) \citd{4r}{5r}. However, this
simple relation does not continue to hold at the full nonlinear level,
since a metric $\widetilde{g}(\widetilde{m},\widetilde{\ell})$ that is
dual to a Taub-NUT metric $g(m,l)$ (in the sense of eq. (\ref{1.2e}))
becomes singular in the limit $\widetilde{\ell}\to 0$ or $m \to 0$
\cite{12r}. Using the modified duality transformation rule for the
Riemann tensor, we find an exact relation between Taub-NUT-(A)dS
metrics 
$\widetilde{g}(\widetilde{m},\widetilde{\ell},\widetilde{\Lambda})$ and
$g(m,\ell,\Lambda)$. Only then we find that a Schwarzschild metric
$\widetilde{g}(\widetilde{m},\widetilde{\ell}\to
0,\widetilde{\Lambda}\to 0)$ in Minkowski space can be obtained as the
dual of a suitable contraction of a Taub-NUT-(A)dS metric $g(m\to
0,\ell\to 0,\Lambda\to -\infty)$ (with $m/\ell$ and $m^3\Lambda$
fixed).

However, the modified duality transformation rule for the Riemann
tensor including a four form field strength is still not sufficient to
allow for dual metrics $\widetilde{g}$ of the form of FRW
cosmologies. The reason is that for such metrics $\widetilde{g}$ the
Ricci tensor is not constant, but the above duality transformation
relates the Ricci tensor of $\widetilde{g}$ to the first Bianchi
identity (the cyclic identity), up to a constant, of the
original Riemann tensor. A solution of this problem consists in a
further modification of the duality transformation: In the space time
described by $g$, we propose to couple matter to gravity in the form of
torsion in the Riemann tensor $R_{abcd}(g)$.

A priori, in the presence of torsion in $R_{abcd}(g)$ it is still
highly non-trivial to generate a Riemann tensor
$\widetilde{R}_{abcd}(\widetilde{g})$ through a duality transformation,
that is torsion free and can be derived from a FRW like metric
$\widetilde{g}$. Nevertheless it turns out that a quite simple ansatz
for the torsion in $R_{abcd}(g)$ -- represented as non-metric
contributions to the connection -- does the job: it suffices to include
torsion in the form of a vector and an axial vector, whose only
nonvanishing components are its time like components $\gamma (t)$ and
$\beta (t)$, respectively. Including in addition a FRW like metric $g$
with a scale factor $a(t)$, one finds that two relations among
$a(t)$, $\gamma (t)$ and $\beta (t)$ are sufficient in order to
generate a dual Riemann tensor that can be derived from a FRW metric
$\widetilde{g}$ with an arbitrary scale factor $\widetilde{a}(t)$. The
non-standard form of the duality relation plays a crucial role to this
end.\par

The subsequent outline of the paper is as follows: In the next section
2 we review the properties of standard (linearized) gravitational
$S$-duality. In section 3 we present a non-standard gravitational
duality rule including a 3-form field. We discuss its consistency at
the linearized level, and derive the result mentioned above: under a
simple assumption flat Minkowski space appears as the dual of (A)dS,
for any value of the de Sitter curvature.\par

In section 4 we generalize this result to Taub-NUT-(A)dS metrics, and
derive the Schwarz\-schild metric as a contraction of a dual
Taub-NUT-AdS metric.\par

In section 5 we consider Riemann tensors with torsion, and derive FRW
cosmologies as duals of theories with torsion. In section 6 we conclude
with a discussion.

\mysection{The Dual of Gravity} 
For most of the paper it will be convenient to work with tensors with
(latin) indices in (flat) tangent space, that are related to tensors
with (greek) indices, as usual, by contractions with a vierbein. For the
Riemann tensor this relation reads
\beq \label{2.1e} R_{abcd} = e^{\ \mu}_a \ e^{\ \nu}_b \ e_c^{\ \rho} \
e_d^{\ \sigma} \ R_{\mu\nu\rho\sigma} \ . \eeq

Let us recall the symmetry properties of $R_{abcd}$:
\beq \label{2.2e} R_{abcd} = - R_{bacd} = -
R_{abdc} = + R_{cdab} \ . \eeq

\noi It satisfies the first Bianchi identity (or cyclic identity)
\beq \label{2.3e} R_{abcd} + R_{acdb} + R_{adbc} = 0 \eeq

\noi and the second Bianchi identity
\beq \label{2.4e} D_{e} R_{abcd} + D_{c} R_{abde} + D_{d} R_{abec} = 0
\ . \eeq

\noi In the vacuum, the equations of motion imply the vanishing of the
Ricci tensor:
\beq \label{2.5e} R_{\phantom{a}b}^{a} \equiv
R_{\phantom{ab}bc}^{ca} = 0  \eeq

\noi where indices are raised and lowered with the flat metric
$\eta_{ab} = diag(-1,\ 1,\ 1,\ 1)$.

A dual Riemann tensor $\widetilde{R}_{abcd}$ is obtained from
$R_{abcd}$ by a contraction with the antisymmetric tensor
$\varepsilon_{abcd}$:

\beq \label{2.6e}  
\widetilde{R}_{abcd} = {1 \over 2} \ \varepsilon_{abef} \
R^{ef}_{\phantom{ab}cd} \qquad \hbox{or} \qquad
\widetilde{R}_{abcd} = {1 \over 2} \ R_{ab}^{\phantom{ab}ef}\
\varepsilon_{efcd}  \ . \eeq

The properties of $\widetilde{R}_{abcd}$ have previously been discussed
in [3--6]. Its first Bianchi identity follows from the
vanishing of the Ricci tensor (\ref{2.5e}) corresponding to $R_{abcd}$.
Its second Bianchi identity {\it at the linearized level} can be
derived from the second Bianchi identity of $R_{abcd}$ ({\it at the
linearized level}) if, again, $R^a_{\phantom{a}b}$ vanishes. Finally
the first Bianchi identity for $R_{abcd}$, eq. (\ref{2.3e}), implies
the vanishing of the dual Ricci tensor.

Its symmetries together with the Bianchi identities are sufficient to
prove that, at the linearized level,
$\widetilde{R}^{lin}_{\mu\nu\rho\sigma}$ can be written in terms of a
dual linearized metric $\widetilde{h}_{\mu\nu}$ \cite{dub} (the
distinction between latin and greek indices is meaningless at the
linearized level) as

\beq \label{2.11e} \widetilde{R}^{lin}_{\mu\nu\rho\sigma} = {1 \over 2}
\left ( \widetilde{h}_{\mu\sigma ,\nu\rho} + \widetilde{h}_{\nu\rho,
\mu\sigma} - \widetilde{h}_{\mu\rho , \nu\sigma} -
\widetilde{h}_{\nu\sigma, \mu \rho} \right ) \ . \eeq

\noi An explicit formula for $\widetilde{h}_{\mu\nu}$ in terms of
$\widetilde{R}_{\mu\nu\rho\sigma}$ is given in \cite{dub} in the
coordinate gauge $x^{\mu} \widetilde{h}_{\mu\nu} = x^{\nu}
\widetilde{h}_{\mu\nu} = 0$:
\beq \label{2.12e} \widetilde{h}_{\mu\nu}(x) = - \int_0^1 dt \int_0^t
dt't' \ x^{\rho} x^{\sigma} \ \widetilde{R}_{\mu\rho\nu\sigma}(t'x) \ .
\eeq

Thus the $S$-dual of linearized gravity can be constructed
explicitly. The validity of the second Bianchi identity (\ref{2.4e}) for
the dual Riemann tensor beyond the linearized level requires, however,
the knowledge of the dual connections which are not yet constructed at
this point, and this problem has no general solution.

\mysection{The S-Dual of Gravity and a 3-Form Field}
The standard gravitational S-duality transformation (\ref{2.6e}) allows
only to relate metrics with vanishing Ricci tensors \citd{3r}{4r}. In
the present section we present a modified duality transformation that
allows to relate metrics whose Ricci tensors $R^a_{\phantom{a} b}$
satisfy
\beq \label{3.1e} R_{\ a}^b \equiv R_{\phantom{ab}bc}^{ca} = \Lambda 
\delta_{\ b}^a \ . \eeq

As stated in the introduction, we assume the presence of a three form
field $A_{abc} = A_{[abc]}$, with field strength $F_{abcd} = 
\partial_{[a} A_{bcd]}$ and equation of motion $\partial^a F_{abcd} =
0$, and study a modified duality transformation that mixes the Riemann
tensor with $F_{abcd}$. As general solution of the equation of motion
of the three form field we can take
\beq \label{3.2e}
F_{abcd} = \Sigma \varepsilon_{abcd} \qquad , \qquad \Sigma = 
\hbox{const.} \ .
\eeq

The proposed generalization of the duality transformation (2.6) reads:
\bminiG{3.3e}
\label{3.3ae}
\widetilde{R}_{abcd} = {1 \over 2} \varepsilon_{abef} \left ( 
R_{\phantom{ef}cd}^{ef} + F^{ef}_{\phantom{ef}cd} \right ) + {1 \over
12} \  \varepsilon_{abcd} R \ ,
\eeeq
\beeq
\label{3.3be}
\widetilde{F}_{abcd} = - {1 \over 12} \ \varepsilon_{abcd} R 
\emini

\noi(or with the reversed order of $\varepsilon$ and $(R + F)$ in
(\ref{3.3ae}), cf. the second of eqs. (2.6)), where $R \equiv
R^{ab}_{\phantom{ab}ba}$.
 
In order to justify eqs. (\ref{3.3e}) we have to show that they imply
an exchange of equations of motion with Bianchi identities, and that a
double duality transformation acts as minus the identity (in space time
with Minkowski signature). 

Let us first discuss the Bianchi identities to be satisfied by the dual
Riemann tensor $\widetilde{R}_{abcd}$ in (\ref{3.3ae}) (the Bianchi
identity for $\widetilde{F}_{abcd}$ is trivial in d=4). Using
(\ref{3.2e}) and the gravitational equation (\ref{3.1e}) it is
straightforeward to show that $\widetilde{R}_{abcd}$ has the symmetry
properties (\ref{2.2e}), and satisfies the first Bianchi identity
(\ref{2.3e}). As in the case of the ``standard" duality relation (2.6),
the validity of the second Bianchi identity (2.4) can only been shown
at the linearized level, where one has to use the linearized second
Bianchi identity for $R_{abcd}$, and the fact that both the Ricci
tensor $R_{\ b}^a$ and $F_{abcd}$ are constant. 

These properties of $\widetilde{R}_{abcd}$  are sufficient to prove
that, at the linearized level, it can again be expressed in terms of a
dual metric $\widetilde{h}_{\mu\nu}$ as in eq. (\ref{2.11e}). \par

Now we turn to the equations of motion satisfied by the dual tensors.
For the dual Ricci tensor one obtains
\beq \label{3.4e} 
\widetilde{R}_{\ b}^a =  3 \Sigma \delta_{\ b}^a
\equiv \widetilde{\Lambda} \delta_{\ b}^a
\eeq

\noi with the help of the first Bianchi identity for $R_{abcd}$, and
eq. (\ref{3.2e}) for $F_{abcd}$.

For the dual four form field strength $\widetilde{F}_{abcd}$ one finds,
from $R = 4 \Lambda$ and eq. (\ref{3.3be}), 
\beq \label{3.5e}
\widetilde{F}_{abcd} = - {1 \over 3}  \Lambda \varepsilon_{abcd} 
\equiv \widetilde{\Sigma} \varepsilon_{abcd}\ ,
\eeq

\noi which is indeed a solution of the dual equations of motion for
$\widetilde{A}_{abc}$.

Equations (\ref{3.4e}) and (\ref{3.5e}) show that in some sense
$A_{abc}$ is dual to the cosmological constant: Up to a factor $3$ the
duality transformations (\ref{3.3e}) lead to an interchange of
$\Sigma$, the parameter characterizing the solution of the equation of
motion of $A_{abc}$, with the cosmological constant $\Lambda$. 

As stated in the introduction, here we have to assume that the duality
transformations (\ref{3.3e}) are accompagnied by duality
transformations acting on the stress energy tensor such that the dual
stress energy tensor is consistent with $\widetilde{\Lambda} =
3\Sigma$. Note that, in the absence of a parent action, we cannot
determine a priori the contribution of $F_{abcd}$ to the stress energy
tensor (or its dual) -- this contribution is not fixed by the
equation of motion $\partial^a F_{abcd} = 0$, but implicitely by the
gravitational equation of motion (\ref{3.4e}).

Finally we have to check whether a double duality transformation
reproduces (minus) the identity. After some calculation one obtains
indeed
\beq \label{3.6e} \widetilde{\widetilde{R}}_{abcd} = - R_{abcd}
\eeq

\noi and
\beq \label{3.7e} \widetilde{\widetilde{F}}_{abcd} = - F_{abcd}
\ . \eeq

Hence the duality transformations (\ref{3.3e}) have all the
desirable properties. As before, however, the validity of a second
Bianchi identity for $\widetilde{R}_{abcd}$ can not be proven beyond
the linearized level.

Let us make a comment on the need to include the three form field
$A_{abc}$ in the duality relation (\ref{3.3ae}). At first sight, we
could have omitted $F_{abcd}$ in (\ref{3.3ae}), and dropped
(\ref{3.3be}): Then $\widetilde{R}_{abcd}$ would still satisfy the
Bianchi identities (at the same linearized level as before), and the
gravitational equations of motion with $\widetilde{R}_{\ b}^a = 0$.
However, then a double duality transformation would reproduce (minus)
the identity on $R_{abcd}$ only iff $\Lambda = 0$. It is the validity
of (\ref{3.6e}) for $\Lambda$ arbitrary that forces us to include 
$F_{abcd}$ in (\ref{3.3ae}).

Now we make the following evident, but important, observation: {\it
Iff} the vev $\Sigma$ of the 3-form field strength (before the duality
transformation) vanishes, the dual Ricci tensor vanishes by virtue of
eq. (\ref{3.4e}), independently from the value $\Lambda$ of the
space-time described by the Riemann tensor $R_{abcd}$ before the
duality transformation. Hence, {\it iff} for some reason we ``see" the
space-time described by the dual Riemann tensor $\widetilde{R}_{abcd}$,
we then see a space-time with vanishing cosmological constant.

As discussed in the introduction, we may identify $\widetilde{R}_{abcd}$
with our ``physical" space-time only if $\widetilde{R}_{abcd}$ can also
describe physically relevant non-trivial configurations as the
Schwarzschild and FRW metrics, and this beyond the linearized
level (such that the full second Bianchi identity (\ref{2.4e}) holds).
The analysis of the action of the generalized duality transformations
(\ref{3.3ae}) on metrics that are suitable generalizations of the
Schwarzschild metric is the subject of the next chapter.

\mysection{Non-standard Duality and Taub-NUT-(A)dS Spaces} 
At the level of linearized standard gravitational S-duality, the
parameters $m$ and $\ell$ of a Taub-NUT metric \citd{10r}{11r} get
interchanged  \citd{4r}{5r}. Hence the NUT parameter $\ell$ can be
interpreted as a ``magnetic" mass. On Taub-NUT spaces the gravitational
S-duality can be extended to the full nonlinear level \cite{12r}, and
this can be generalized to Taub-NUT-(A)dS spaces in the case of the
non-standard gravitational S-duality (\ref{3.3e}) \cite{12r}. At the
nonlinear level, however, the relations between the original parameters
$m$, $\ell$, and the parameters $\widetilde{m}$, $\widetilde{\ell}$
caracterizing the dual configuration, are somewhat more involved (see
below).

The Taub-NUT-(A)dS metric can be written as the following
generalization of the Taub-NUT metric \cite{stef}:
\beq \label{4.1e} ds^2 = - f^2(r) \left ( dt + 4 \ell \sin^2 {\theta
\over 2} d\phi \right )^2 + f^{-2}(r) dr^2 + \left ( r^2 + \ell^2
\right ) \left ( d\theta^2 + \sin^2 \theta d\phi^2 \right ) \eeq

\noi with
\beq \label{4.2e} f^2(r) = 1 -{2\left ( mr + \ell^2 \right ) -
\Lambda \left ( {1 \over 3} r^4 + 2 \ell^2 r^2 - \ell^4 \right )
\over r^2 + \ell^2} \ . \eeq

\noi The non-vanishing components of the Riemann tensor are
\bea \label{4.3e} 
&&R_{0101} = - 2 \left ( 1 + {4 \over 3} \Lambda \ell^2 \right )
A_{\bar{m}, \ell} (r)+{1 \over 3} \Lambda \nn \\
&&R_{0202} = R_{0303} =  \left ( 1 + {4 \over 3} \Lambda \ell^2 \right )
A_{\bar{m}, \ell} (r)+{1 \over 3} \Lambda\nn \\
&&R_{1212} = R_{1313} = - \left ( 1 + {4 \over 3} \Lambda \ell^2 
\right ) A_{\bar{m}, \ell} (r)-{1 \over 3} \Lambda\nn \\ 
&&R_{2323} = 2 \left ( 1 + {4 \over 3} \Lambda \ell^2 
\right ) A_{\bar{m}, \ell} (r)-{1 \over 3} \Lambda \nn \\ 
&&R_{0312} = - R_{0213} = \left ( 1 + {4 \over 3} 
\Lambda \ell^2 \right ) D_{\bar{m}, \ell} (r) \nn \\
&&R_{0123} = - 2 \left ( 1 + {4 \over 3} 
\Lambda \ell^2 \right ) D_{\bar{m}, \ell} (r)\ , \eea

\noi where $A_{\bar{m}, \ell}$ and $D_{\bar{m}, \ell}$ are given by
\bea \label{4.4e} &&A_{\bar{m}, \ell }(r) ={\bar{m} r^3 + 3 \ell^2 r^2
- 3 \bar{m} \ell^2 r - \ell^4 \over (r^2 + \ell^2)^3} \ ,  \nn \\
&&D_{\bar{m}, \ell }(r) ={- \ell r^3 + 3 \ell \bar{m}r^2 + 3 r \ell^3 -
\bar{m}\ell^3  \over (r^2 + \ell^2)^3}  \eea

\noi and
\beq \label{4.5e} \bar{m} = m \left ( 1 + {4 \over 3} \Lambda
\ell^2 \right )^{-1} \ . \eeq

Constructing the components of the dual Riemann tensor from eq.
(\ref{3.3ae}) one obtains contributions from the terms
$\sim F_{abcd}$ and $\sim R$. One finds
\bea \label{4.6e}
&&\widetilde{R}_{0101} = - 2 \left ( 1 + {4 \over 3} 
\Lambda \ell^2 \right ) D_{\bar{m}, \ell} (r) + \Sigma \nn \\
&&\widetilde{R}_{0202} = \widetilde{R}_{0303} = \left ( 1 + {4 \over
3}  \Lambda \ell^2 \right ) D_{\bar{m}, \ell} (r) + \Sigma \nn \\
&&\widetilde{R}_{1212} = \widetilde{R}_{1313} = - \left ( 1 + {4 \over
3}  \Lambda \ell^2 \right ) D_{\bar{m}, \ell} (r) - \Sigma \nn \\
&&\widetilde{R}_{2323} = 2 \left ( 1 + {4 \over 3} 
\Lambda \ell^2 \right ) D_{\bar{m}, \ell} (r) - \Sigma \nn \\
&&\widetilde{R}_{0312} = \widetilde{R}_{0213} = - \left ( 1 + {4 \over
3} \Lambda \ell^2  \right ) A_{\bar{m}, \ell} (r) \nn \\
&&\widetilde{R}_{0123} = 2 \left ( 1 + {4 \over 3} \Lambda \ell^2 
\right ) A_{\bar{m}, \ell} (r) \ .
\eea

The following properties of the functions $A_{{m}, \ell}$, $D_{{m},
\ell}$ are helpful in order to find a metric that reproduces the
Riemann tensor (\ref{4.6e}): If one defines
\beq \label{4.7e} m' = - {\ell^2 \over m} \eeq

\noi one has
\bea \label{4.8e} &&A_{m',\ell}(r) =  \ {\ell \over m} \ D_{m,
\ell}(r) \ , \nn \\ &&D_{m',\ell}(r) = - {\ell \over m} \ A_{m, \ell}(r)
\ . \eea

Note that, at the level of linearized gravity, we could replace
$A_{m, \ell}(r)$ and $D_{m, \ell}(r)$ in (\ref{4.4e}) by their
asymptotic forms for $r \to \infty$. Then we would obtain the simple
relation $A_{m, \ell}(r) = - D_{\ell, m}(r)$. This
simple relation does not survive in full non-linear gravity.

Using the relations (\ref{4.8e}) and the definition (\ref{4.5e}) of 
$\bar{m}$ one finds that the following metric reproduces all components
of $\widetilde{R}_{abcd}$:
\beq
\label{4.9e}
\widetilde{ds}^2 = - {\ell \over m - 4 \Sigma \ell^3} \left  \{
\widehat{f}^{\, 2}(r) \left ( dt + 4 \ell \sin^2 {\theta \over 2} d
\phi \right )^2 - \left [ \widehat{f}^{-2}(r) dr^{2} 
+ \left ( r^{2} + \ell^2 \right ) \left ( d \theta^2
+ \sin^2 \theta d\phi^2 \right ) \right ] \right \}
\eeq

\noi with
\beq \label{4.10e} \widehat{f}^{\, 2}(r) = 1 + {-2 \ell^2 \left ( m - 4
\Sigma \ell^3 - r\left ( 1 + {4 \over 3} \Lambda \ell^2 \right ) \right
) + 3 \Sigma \ell \left ( {1 \over 3}r^4 + 2 r^2 \ell^2 - \ell^4 \right
) \over \left ( m - 4 \Sigma \ell^3 \right ) \left ( r^2 + \ell^2
\right )} \ . \eeq

In order to bring this metric into the same form as in (\ref{4.1e}) one
has to rescale the coordinates as
\beq \label{4.11e} t = \sqrt{{m - 4 \Sigma \ell^3 \over \ell}} \
t' \qquad , \qquad r = \sqrt{{m - 4 \Sigma \ell^3 \over \ell}} \
r' \ , \eeq

\noi and to define the dual parameters
\bea \label{4.12e} &&\widetilde{m} = - \left ( 1 + {4 \over 3}  \Lambda
\ell^2 \right )\left ( m - 4 \Sigma \ell^3 \right )^{-3/2} \ell^{5/2}
\quad , 
\quad \widetilde{\ell} = \ell^{3/2}
\left ( m - 4 \Sigma \ell^2 \right )^{-{1 \over 2}} \ ,\nn \\
&&\widetilde{\Lambda} = 3 \Sigma \ . \eea

This allows to write the dual metric again (up to an overall sign) in
the form (\ref{4.1e}) with
\beq \label{4.13e} \widetilde{f}^2 (r') = 1 - {2\left ( \widetilde{m}
r' + \widetilde{\ell}^2 \right ) - \widetilde{\Lambda} \left ( {1 \over
3}r^{'4} + 2 r^{'2} \widetilde{\ell}^2 - \widetilde{\ell}^4 \right )
\over r^{'2} + \widetilde{\ell}^2} \ . \eeq

Thus the metric dual to a Taub-NUT-(A)dS metric is again of the
Taub-NUT-(A)dS form. Let us now assume that the vev $\Sigma$ of the
3-form field strength vanishes. Then the expressions (\ref{4.12e}) for
the dual parameters collapse to
\beq \label{4.14e} 
\widetilde{m} = - \left( 1 + {4 \over 3}  \Lambda \ell^2 \right)
m^{-3/2} \ell^{5/2} \ , 
\ \widetilde{\ell} = \ell^{3/2} m^{-{1 \over 2}} \ ,\
\widetilde{\Lambda} = 0 \ . \eeq

Now, as in section 3, the dual cosmological constant vanishes, but,
somewhat disturbingly, the dual NUT parameter $\widetilde{\ell}$
does not vanish for $m \to 0$ (in contrast to its behaviour in
linearized gravity). However, a vanishing dual NUT parameter -- as
required for a dual Schwarzschild metric -- can be obtained in the
following limit:
\beq \label{4.15e}
\Lambda \to -\infty,\quad m,\ \ell \to 0,\quad m/\ell = k = const.
\eeq

\noi It turns out that the constant k can be absorbed into a rescaling
of the coordinates and be chosen as $k = 1$. Then, in addition, we
require
\beq \label{4.16e}
-{4\over 3} m^3 \Lambda = \widetilde{m} = const.
\eeq
\noi when taking the limits (\ref{4.15e}). Now, since $\widetilde{\ell}
\to 0$, the dual metric (as described by $\widetilde{f}^2 (r')$ in
(\ref{4.13e}) with $\widetilde{\ell} = \widetilde{\Lambda} = 0$)
coincides with the Schwarzschild metric with mass $\widetilde{m}$.

Note that during the above contraction of the original metric we have
kept the coordinates $r$, $t$ constant, which is a coordinate dependent
statement. As usual in the case of contractions, coordinates have
eventually be rescaled after a parameter dependent general coordinate
transfromation.

Hence we have obtained the desired result: the Schwarzschild metric can
be obtained as the dual of a contracted Taub-NUT-AdS metric. This
result would not have been possible in the absence of an "original"
cosmological constant $\Lambda$, and using the standard gravitational
S-duality transformation. Note that although the original metric (and
the components of the original Riemann tensor) diverge in the above
limit (\ref{4.15e}), these infinities cancel in the non-standard
expression (\ref{3.3ae}) for the dual Riemann tensor which is what
makes this result possible.

\mysection{FRW Cosmologies as Duals of Gravity with
\newline\noindent
Torsion} \hspace*{\parindent}
In this section we investigate whether a dual Riemann tensor
$\widetilde{R}_{abcd}$, obtained through a non-standard duality
transformation of the type (3.3a), can be identified with a Riemann
tensor describing FRW cosmologies. FRW cosmologies correspond to a
metric

\beq
\label{5.1e}
d\widetilde{s}^2 = - dt^2 + \widetilde{a}^2(t)\ d\vec{x}^2
\eeq
\noi where, of course, $\widetilde{a}(t)$ depends on the properties 
of the matter to which the Einstein tensor couples.\par

Defining

\beq
\label{5.2e}
\widetilde{a}(t) = e^{\widetilde{\alpha}(t)}
\eeq

\noi the only nonvanishing components of the Riemann tensor 
$\widetilde{R}_{abcd}$ are

\bminiG{5.3e}
\label{5.3ae}
\widetilde{R}_{ijij} =  \dot{\widetilde{\alpha}}^2\ \hbox{(no sum 
over $i$, $j$)}\ ,
\eeeq
\beeq
  \label{5.3be}
  \widetilde{R}_{i0i0} =  - \dot{\widetilde{\alpha}}^2 - 
\ddot{\widetilde{\alpha}}\ ,
\emini

\noi and the nonvanishing components of the Ricci tensor are

\bminiG{5.4e}
\label{5.4ae}
\widetilde{R}_{ii} = - 3 \dot{\widetilde{\alpha}}^2- 
\ddot{\widetilde{\alpha}}\ ,
\eeeq
\beeq
  \label{5.4be}
  \widetilde{R}_{00} =  3 \dot{\widetilde{\alpha}}^2 + 3 
\ddot{\widetilde{\alpha}}\ ,
\emini

\noi where dots denote time derivatives.\par

However, the duality transformation (3.3a) allows only for Ricci
tensors $\widetilde{R}_{ab} = 3 \Sigma \eta_{ab}$ (see (3.4)) with
$\dot{\Sigma} = 0$ from the equation of motion (3.2) for the 3-form
field. Hence the duality transformation (3.3a) has to be modified by
additional terms corresponding to contributions from matter in the
``original'' version of the theory before the duality transformation.
\par

The most elegant way to do this is to replace the Riemann tensor
$R_{abcd}$ on the right-hand side  of (3.3a) by a Riemann tensor
including torsion \cite{13r}. Our corresponding conventions are as
follows: the Riemann tensor is written as

\beq
\label{5.5e}
R_{\phantom{\sigma}\mu\nu\rho}^{\sigma} =
\Gamma^{\sigma}_{\phantom{\sigma}\mu \rho , \nu} - 
\Gamma_{\phantom{\sigma}\mu \nu , \rho}^{\sigma} +
\Gamma_{\phantom{\sigma}\beta \nu}^{\sigma} 
\Gamma_{\phantom{\sigma}\mu\rho}^{\beta} -
\Gamma_{\phantom{\sigma}\beta \rho}^{\sigma} 
\Gamma_{\phantom{\sigma}\mu\nu}^{\beta}
\eeq

\noi where the connection is decomposed as

\beq
\label{5.6e}
\Gamma_{\phantom{\sigma}\mu\nu}^{\sigma} =\
^M\Gamma_{\phantom{\sigma}\mu\nu}^{\sigma} + 
\widehat{\Gamma}_{\phantom{\sigma}\mu\nu}^{\sigma} \ .
\eeq

\noi Here $^{M}\Gamma_{\phantom{\sigma}\mu\nu}^{\sigma}$ is the
standard connection  constructed from the metric $g_{\mu\nu}$, and 
$\widehat{\Gamma}_{\phantom{\sigma}\mu\nu}^{\sigma}$ represents
torsion. Requiring  $g_{\mu\nu ; \rho} = 0$ (where the covariant
derivative is defined  with the full connection
$\Gamma_{\phantom{\sigma}\mu\nu}^{\sigma}$) implies

\beq
\label{5.7e}
\widehat{\Gamma}_{\sigma \mu\nu} = \widehat{\Gamma}_{[\sigma \mu]\nu}
\ ,
\eeq

\noi where indices are raised and lowered with the metric
$g_{\mu\nu}$. Assuming eq. (\ref{5.7e}), $\widehat{\Gamma}_{\sigma
\mu\nu}$ can be  decomposed with respect to the Lorentz group as 
\citm{13r}{15r}

\beq
\label{5.8e}
\widehat{\Gamma}_{\sigma \mu\nu}  = \widehat{\Gamma}_{\sigma 
\mu\nu}^V + \widehat{\Gamma}_{\sigma \mu\nu}^A + 
\widehat{\Gamma}_{\sigma \mu\nu}^T \ .
\eeq

\noi Here $\widehat{\Gamma}_{\sigma \mu\nu}^V$ is proportional to a 
vector $V_{\mu}$,

\beq
\label{5.9e}
\widehat{\Gamma}_{\sigma \mu\nu}^V = V_{[\sigma} \ g_{\mu]\nu} \ ,
\eeq

\noi $\widehat{\Gamma}_{\sigma \mu\nu}^A$ is totally antisymmetric and
proportional to an axial vector $A_{\mu}$,

\beq
\label{5.10e}
  \widehat{\Gamma}_{\sigma \mu\nu}^A = \varepsilon_{\sigma \mu\nu 
\rho} \ A^{\rho}
\eeq

\noi and $\widehat{\Gamma}_{\sigma \mu\nu}^T$ is traceless. For our
subsequent purposes -- the discussion of cosmologies -- it suffices to
confine ourselves to torsion of the type $\widehat{\Gamma}_{\sigma
\mu\nu}^V$ and $\widehat{\Gamma}_{\sigma \mu\nu}^A$ \cite{14r}.
Moreover, according to the symmetries associated to the cosmological
principle (isotropy and homogeneity), only the time (zero) components
of $V_{\mu}$ and $A_{\mu}$ are assumed to be nonvanishing and to depend
on $t$ only. \par

First, we make an ansatz for the metric $g$ analogous to eq.
(\ref{5.1e}),

\beq
\label{5.11e}
ds^2 = - dt^2 + a^2(t) \ d\vec{x}^2 \ .
\eeq

Then it turns out to be  convenient to parametrize the nonvanishing 
components of $\widehat{\Gamma}_{\sigma \mu\nu}^V$ as

\beq
\label{5.12e}
\widehat{\Gamma}_{0ij}^V = - \widehat{\Gamma}_{i0j}^V = \delta_{ij}\ 
a^2(t) \ \gamma (t) \ ,
\eeq

\noi and the nonvanishing components of $\widehat{\Gamma}_{\sigma
\mu\nu}^A$ as

\beq
\label{5.13e}
\widehat{\Gamma}_{ijk}^A = \varepsilon_{ijk}\ a^3(t)\ \beta (t) \ .
\eeq

Assuming the existence of an action $S$, the full connection
$\Gamma_{\phantom{\sigma}\mu\nu}^{\sigma}$ and the  metric $g_{\mu\nu}$
are determined by varying

\beq
\label{5.14e}
S = \int d^4x \left \{ {1 \over 2} \sqrt{-g}\ g^{\mu\nu} \ 
R^{\sigma}_{\phantom{\sigma}\mu\nu\sigma} (\Gamma ) + {\cal L}_m (g,
\Gamma , \cdots  ) \right \}
\eeq

\noi both with respect to $g_{\mu\nu}$ and
$\Gamma_{\phantom{\sigma}\mu\nu}^{\sigma}$ \citd{13r}{14r} (where
${\cal L}_m$ is the Lagrangian of matter fields). In the context of
cosmology suitable averages over the matter fields are performed, and
the resulting equations can be expressed in terms of an ``effective''
density (depending on the torsion), an ``effective'' pressure and
sources for torsion, whose unknown properties allow to treat the
functions $\gamma (t)$, $\beta (t)$ in eqs. (\ref{5.12e}) and
(\ref{5.13e}) as additional arbitrary parameters \citd{14r}{15r}. \par

Here we are not interested in the dynamics that fixes $a(t)$, $\gamma
(t)$ and $\beta (t)$, but rather in the following problem: Given the
three above functions, we can construct the Riemann tensor (\ref{5.5e})
or its version $R_{abcd}$ according to (2.1). Then we can find its dual
according to eq. (3.3a) and ask, whether the components of
$\widetilde{R}_{abcd}$ can be of the form of eqs. (\ref{5.3e}) such
that they describe standard -- torsionless -- FRW cosmologies. \par

This is a highly nontrivial question, since
$R_{\phantom{\sigma}\mu\nu\rho}^{\sigma}$ has none of the properties
(2.2) -- (2.4) due to the presence of torsion. (Of course, we
introduced torsion in order to avoid  the cyclic identity (2.3) which
implies the vanishing of $\widetilde{R}_{\phantom{\sigma}b}^a$, but now
it can well be impossible to satisfy all of the constraints (2.2),
(2.4) for $\widetilde{R}_{abcd}$.)\par

First, $R_{\phantom{\sigma}\mu\nu\rho}^{\sigma}$ as obtained from eq.
(\ref{5.5e}) with $\Gamma_{\phantom{\sigma}\mu\nu}^{\sigma}$ as in eq.
(\ref{5.6e}), a metric as in (\ref{5.11e}) and
$\widehat{\Gamma}^{\sigma}_{\phantom{\sigma}\mu\nu}$ given by the sum
of eqs. (\ref{5.12e}) and (\ref{5.13e}), is no longer symmetric:

\beq
\label{5.15e}
R_{\sigma \mu \nu \rho} \not= R_{\nu\rho \sigma \mu} \ \hbox{(in
general)} \ .
\eeq

Consequently the result for $\widetilde{R}_{abcd}$ depends on whether,
in the duality transformation, one contracts $\varepsilon_{abcd}$ to
the left of $R_{abcd}$ (as in eq. (3.3a)),  to the right of $R_{abcd}$,
or whether one employs a left-right symmetric definition of the duality
transformation. Below we will treat all possible cases. \par

Recall that, originally, the duality transformation (3.3a) leads to flat
Minkowski space (described by $\widetilde{R}_{abcd}$) if the field
strength $F_{abcd}$ vanishes, regardless of the cosmological constant
(curvature of (A)dS space) described by $R_{abcd}$. We will continue
to work with the assumption of vanishing $F_{abcd}$. However, in order
to treat the different possible duality transformations simultaneously,
we generalize (3.3a) as

\beq
\label{5.16e}
\widetilde{R}_{abcd} = {1 \over 4} \left [ (1 + e)\varepsilon_{abef}  \
R^{ef}_{\phantom{\sigma\sigma}cd} + (1-e)
R_{ab}^{\phantom{\sigma\sigma}ef}\ \varepsilon_{efcd} \right ] + {1 
\over 12}\ \varepsilon_{abcd}\ R \ .
\eeq

We have dropped the terms $\sim F_{abcd}$, but the parameter $e$ allows
to interpolate between \par

i) left duality ($e = 1$)\par

ii) right duality ($e = -1$)\par

iii) left-right symmetric duality ($e = 0$). \par

Our results concerning the properties of $\widetilde{R}_{abcd}$ are as
follows: First, $\widetilde{R}_{abcd}$ satisfies all of the symmetry
properties (2.2) (where the last one is nontrivial) if and only if the
three functions $\alpha$, $\beta$ and $\gamma$ satisfy

\beq
\label{5.17e}
e\left ( \beta^2 - \gamma^2 + \dot{\gamma} - \dot{\alpha}\gamma + 
\ddot{\alpha} \right ) = 0 \ .
\eeq

Second, $\widetilde{R}_{abcd}$ satisfies the cyclic identity (2.3) if 
and only if

\beq
\label{5.18e}
e\left ( \beta^2 - \gamma^2 + \dot{\gamma} - \dot{\alpha}\gamma + 
\ddot{\alpha} \right ) = 0 \ .
\eeq

The fact that eqs. (\ref{5.17e}) and (\ref{5.18e}) coincide is not
trivial; the presence of the last term $\sim \varepsilon_{abcd} R$ in
(\ref{5.16e}) is crucial to this end. Then it is quite remarkable that
a very large number of constraints is satisfied simultaneously once
either $e = 0$, or one particular relation between $\alpha (t)$, $\beta
(t)$ and $\gamma (t)$ is satisfied. \par

In terms of the {\it original} Ricci tensor $R_{\phantom{\sigma}b}^a$
(before the duality transformation) the relation (\ref{5.17e})
corresponds to $R_{ii} = - R_{00}$ (no sum over $i$), or

\beq
\label{5.19e}
R_{ab} = \lambda (t) \ \eta_{ab}
\eeq

\noi for some function

\beq
\label{5.20e}
\lambda (t) = - 3 \left ( \dot{\alpha}^2 + \ddot{\alpha} + 
\dot{\alpha}\gamma + \dot{\gamma}\right ) \ .
\eeq

Once eq. (\ref{5.17e}) holds, the only nonvanishing components of 
$\widetilde{R}_{abcd}$ are

\bminiG{5.21e}
\label{5.21ae}
\widetilde{R}_{ijij} = {1 \over 2} \ (1+e) \dot{\beta} + (1 - e)\beta 
\gamma + {(3-e)\over 2} \dot{\alpha}\beta \ ,
\eeeq
\beeq
  \label{5.21be}
  \widetilde{R}_{i0i0} =  - {1 \over 2} \ (1-e) \dot{\beta} - (1 + 
e)\beta \gamma - {(3-e)\over 2} \dot{\alpha}\beta \ .
\emini

Instead of investigating the validity of the second Bianchi identity
(2.4) for $\widetilde{R}_{abcd}$, we can study directly whether eqs.
(\ref{5.21e}) can coincide with eqs. (\ref{5.3e}), that would describe
a FRW cosmology in terms of $\widetilde{R}_{abcd}$.\par

First, we find that for $e= 0$ (left-right symmetric duality) the two
expressions (\ref{5.21ae}) and (\ref{5.21be}) coincide up to a sign,
which implies, from eqs. (\ref{5.3e}), that $\ddot{\widetilde{\alpha}}
= 0$ or

\beq
\label{5.22e}
\dot{\widetilde{\alpha}} = \ {\rm const.}\ = \pm H \ .
\eeq

Hence, the left-right symmetric dual of a cosmology with torsion
corresponds necessarily to (A)dS, what is not general enough for our
purposes. \par

On the other hand, for both cases $e = \pm 1$ we can describe {\it any}
cosmology $\widetilde{\alpha}(t)$ if, in addition to eq. (\ref{5.17e}),
the three functions $\alpha$, $\beta$ and $\gamma$ satisfy the
following relation: \par

 From eqs. (\ref{5.3e}) one can derive

\beq
\label{5.23e}
\widetilde{R}_{ijij} + \widetilde{R}_{i0i0} = {d \over dt} \left ( 
\widetilde{R}_{ijij}\right )^{1/2} \ ,
\eeq

\noi and -- after the use of eqs. (\ref{5.21e}) -- the satisfaction of
the corresponding additional differential equation between $\alpha
(t)$, $\beta (t)$ and $\gamma (t)$ is sufficient in order to be able to
write eqs. (\ref{5.21e}) in the form of eqs. (\ref{5.3e}) with 
$\dot{\widetilde{\alpha}} =  \left (\widetilde{R}_{ijij}\right )^{1/2}$
and $\left (\widetilde{R}_{ijij}\right )^{1/2}$ as in eq. (5.21a).
Since, for $\widetilde{\alpha} (t)$ given, we have only two equations
(\ref{5.17e}) and (\ref{5.23e}) to satisfy, the remaining freedom
allows to describe any cosmology $\widetilde{\alpha} (t)$ with the help
of suitable functions $\alpha (t)$, $\beta (t)$ and $\gamma (t)$.\par

Generally, an explicit solution of the corresponding system of
differential equations (with $\widetilde{\alpha} (t)$ given) is very
difficult to impossible. However, FRW cosmologies corresponding to a
relativistic fluid with a simple equation of state, $p = w \rho$, give
rise to logarithmic scale factors $\widetilde{\alpha}(t)$ with

\beq
\label{5.24e}
\dot{\widetilde{\alpha}} (t) = {2 \over 3}\ (1 + w) t^{-1}\ .
\eeq

\noi In this case all required relations can be satisfied by a simple
ansatz

\bea
\label{5.25e}
&&\dot{\alpha} (t) = a_0\ t^{-1} \nn \\
&&\beta (t)= b_0\ t^{-1} \nn \\
&&\gamma (t) = g_0 \ t^{-1}
\eea

\noi and the ($e$-dependent) solution of a non-linear algebraic system
of 3 equations for the 3 constants $a_0$, $b_0$ and $g_0$.\par

The main result of the present section is, however, the statement made
already above: Given a duality transformation of the form of eq.
(\ref{5.16e}), with $e = \pm 1$, we can obtain any FRW-like Riemann
tensor $\widetilde{R}_{abcd}$ as the dual of an ``original'' theory
with torsion of the form in eqs. (\ref{5.12e}) and (\ref{5.13e}), for
suitable functions $\alpha (t)$, $\beta (t)$ and $\gamma (t)$. The fact
that we manage to satisfy all symmetry conditions and Bianchi
identities for $\widetilde{R}_{abcd}$ simultaneously is highly
nontrivial, and depends on the last term in eq. (3.3a) which can be
considered as a remnant of the duality transformation including the
3-form field (although its field strength has finally been set to
zero).\par

Except for the relation (\ref{5.19e}) we have not been able to express
the required relations between $\alpha (t)$, $\beta (t)$ and $\gamma
(t)$ in terms of dynamical principles of the original theory with
torsion. If these relations do not hold (exactly), the resulting dual
Riemann tensor $\widetilde{R}_{abcd}$ corresponds again to a cosmology
with torsion, a possibility that has been investigated, e.g., in
\citd{14r}{15r}.

\mysection{Discussion} \hspace*{\parindent} Above we considered
generalizations of gravitational $S$-duality. Including a 3-form field
$A_{abc}$, we obtained the particularly interesting result that the
dual cosmological constant vanishes independently of the value of the
``original'' cosmological constant, if the corresponding field strength
vanishes. This motivated us to investigate under which conditions
phenomenologically relevant metrics $\widetilde{g}_{\mu\nu}$ can be
obtained through gravitational $S$-duality transformations. \par

We found that the Schwarzschild metric can be obtained as the dual of a
contracted Taub-NUT-AdS metric. The necessity to perform such a
contraction can be considered as unsatisfactory, but otherwise one is
left with a non-vanishing NUT parameter $\widetilde{\ell}$ in the dual
(supposedly physical) metric. The physics and phenomenology of
non-vanishing NUT parameters has been studied in \cite{16r}. However,
non-vanishing NUT parameters give rise to closed timelike curves, if
one insists on the completeness of the metric \cite{11r}, which seems to
rule out such a possibility. But, for tiny NUT parameters $\ell$, the
argument assumes completeness of the metric at tiny (timelike)
distances. Assuming a modification of gravity (UV regularization) at
small distances, this problem may disappear. For instance, lattice
gauge theories contain Dirac monopoles (whose Dirac string ``escapes''
through the space between the lattice sites). It seems to be a logical
possibility that lattice regularized theories of gravity contain
equally configurations with ``magnetic'' masses (corresponding to NUT
parameters $\ell$), without the above problem of closed timelike
curves. Then we could possibly live with small nonvanishing NUT charges
$\ell$ (up to phenomenological constraints \cite{16r}), and the
contraction performed at the end of chapter 4 does not have to be
pushed to its singular limit.\par

In chapter 5 we obtained FRW-like metrics as duals of theories with
torsion. Clearly, the corresponding Riemann tensor (\ref{5.5e}), given
the decomposition (\ref{5.6e}) of the connection, can always be written
as

\beq \label{6.1e} R_{\phantom{\sigma}\mu\nu\rho}^{\sigma} =\
^{M}R_{\phantom{\sigma}\mu\nu\rho}^{\sigma} +
\widehat{R}_{\phantom{\sigma}\mu\nu\rho}^{\sigma} \eeq

\noi where $^{M}R_{\phantom{\sigma}\mu\nu\rho}^{\sigma}$ depends on the
metric $g_{\mu\nu}$ only, and
$\widehat{R}_{\phantom{\sigma}\mu\nu\rho}^{\sigma}$ depends on the
contribution of torsion to the connection. Thus the presence of torsion
in ${R}_{\phantom{\sigma}\mu\nu\rho}^{\sigma}$ -- appearing on the
right-hand side of the duality transformation (\ref{5.16e}) -- can
equally be interpreted as another generalization of the original
duality transformation rule (3.4a) in the form of adding more matter
(torsion) dependent terms to its right-hand side. However, here matter
is not represented in the form of fields, but in the form of components
of the torsion $\widehat{\Gamma}_{\phantom{\sigma}\mu\nu}^{\sigma}$,
that are treated as effective densities as it is the case for $\rho$
and the pressure $p$ in FRW cosmologies with matter in the form of a
relativistic liquid.\par

Clearly, an application of the present results on metrics (that are
related by duality) to the cosmological constant problem relies on the
possibility to extend the duality transformations in the gravitational
sector consistently to a complete theory including matter: The result
$\widetilde{\Lambda} = 3\Sigma$ in section 3 has to be consistent with
the properties of the dual stress energy tensor $\widetilde{T}_{ab}$. We
remark, however, that the particular case $\widetilde{T}_{ab}=0$ (in the
vacuum) could correspond to a situation where the coupling of matter to
the dual metric is only describable in terms of point like matter, but
not (in a local way) in terms of matter in the form of fields. Such a
situation would be similar to the coupling of magnetic monopoles to an
abelian gauge field $A$, or to the coupling of electric monopoles to its
dual $\widetilde{A}$.

Admittedly such a possibility seems far-fetched, but we recall that
tests of general relativity (or its Newtonian limit) involve only the
coupling of point like matter to gravity, down to length scales of $\sim
1$ mm. (A notable exception is the electromagnetic field, which has to
couple to the dual metric $\widetilde{g}$ within this scenario, such
that light propagates along geodesics corresponding to $\widetilde{g}$.)

In the absence of a duality transformation including matter (that may
well be non local, and may even require to review our present concepts
of space time) we confined ourselves to a ``bottom-up'' approach, in
the sense that we studied macroscopic configurations of the metric
(verified at distances $\gsim$ 1 mm), where matter is represented
either as a point like source at the center of the Schwarzschild
solution, or as effective densities in the case of cosmological
solutions.\par

The fact that both phenomenologically relevant metrics can be obtained,
under suitable assumptions, as $S$-duals indicates that our observed
space time could possibly be identified with the dual of some
underlying metric.

\newpage 

\def\labelenumi{[\arabic{enumi}]} 
\noindent {\large\bf References} 
\ben 

\item\label{west} P.~C.~West, Class.\ Quant.\ Grav.\  {\bf 18} (2001)
4443 [arXiv:hep-th/0104081]; Class.\ Quant.\ Grav.\  {\bf 20}
(2003) 2393 [arXiv:hep-th/0212291].

\item\label{3r} J. Nieto, Phys. Lett. {\bf A262} (1999) 274.

\item\label{4r} C. Hull, Nucl. Phys. {\bf B583} (2000) 237; JHEP {\bf
0012} (2000) 007 [arXiv:hep-th/0011215]; JHEP {\bf 0109} (2001) 27
[arXiv:hep-th/0107149];\\ 
X.~Bekaert and N.~Boulanger, Commun.\ Math.\ Phys.\  {\bf 245} (2004) 27
\\ {}[arXiv:hep-th/0208058];\\
P.~de Medeiros and C.~Hull, Commun.\ Math.\
Phys.\ {\bf 235} (2003) 255 \\ {}[arXiv:hep-th/0208155].

\item\label{5r} H.~Casini, R.~Montemayor and L.~F.~Urrutia, Phys.\
Rev.\ D {\bf 68} (2003) 065011 [arXiv: hep-th/0304228].

\item\label{nurm} A.~J.~Nurmagambetov, ``Duality-symmetric gravity and
supergravity: Testing the PST approach,'' arXiv:hep-th/0407116.

\item\label{henn}
M.~Henneaux and C.~Teitelboim,
``Duality in linearized gravity,''
arXiv:gr-qc/0408101.

\item\label{Zwanziger:1970hk}
D.~Zwanziger,
Phys.\ Rev.\ D {\bf 3} (1971) 880.

\item\label{Schwarz:1993vs}
S.~Deser and C.~Teitelboim, Phys.\ Rev.\ D {\bf 13} (1976) 1592;\\
J.~H.~Schwarz and A.~Sen,
Nucl.\ Phys.\ B {\bf 411} (1994) 35
[arXiv:hep-th/9304154].

\item\label{Witten:1995gf}
E.~Witten,
Selecta Math.\  {\bf 1} (1995) 383
[arXiv:hep-th/9505186].

\item\label{Pasti}
P.~Pasti, D.~P.~Sorokin and M.~Tonin,
Phys.\ Lett.\ B {\bf 352} (1995) 59 [arXiv:hep-th/9503182], 
Phys.\ Rev.\ D {\bf 52} (1995) 4277 [arXiv:hep-th/9506109].

\item\label{kras} A. Krasi\'nski, {\it Inhomogeneous Cosmological 
Models}, Cambridge University Press, Cambridge (UK), 1997.

\item\label{beka} X.~Bekaert, N.~Boulanger and S.~Cnockaert,
``No self-interaction for two-column massless fields,''
arXiv:hep-th/0407102;\\
X.~Bekaert, N.~Boulanger and M.~Henneaux, Phys.\ Rev.\ D {\bf 67}
(2003) 044010 [arXiv:hep-th/0210278].

\item\label{crem}
E.~Cremmer, B.~Julia, H.~Lu and C.~N.~Pope,
Nucl.\ Phys.\ B {\bf 523} (1998) 73 [arXiv:hep-th/9710119], 
Nucl.\ Phys.\ B {\bf 535} (1998) 242 [arXiv:hep-th/9806106].

\item\label{6r} A. Aurilia, H. Nicolai, P. Townsend, Nucl. Phys. {\bf
B176} (1980) 509; \\ M. Duff, P. van Nieuwenhuizen, Phys. Lett. {\bf
B94} (1980) 179; \\ S. Hawking, Phys. Lett. {\bf B134} (1984) 403;\\
J. Brown, C. Teitelboim, Phys. Lett. {\bf B195} (1987) 177; \\ M.
Duff, Phys. Lett. {\bf B226} (1989) 36;\\
A.~Gomberoff, M.~Henneaux, C.~Teitelboim and F.~Wilczek,
Phys.\ Rev.\ D {\bf 69} (2004) 083520 [arXiv:hep-th/0311011].

\item\label{7r} H.~Nishino and S.~Rajpoot, ``Hodge duality and
cosmological constant,'' arXiv:hep-th/0404088.

\item\label{12r} U.~Ellwanger, ``Gravitational S-duality realized on
NUT-Schwarzschild and NUT-de Sitter metrics,'' arXiv:hep-th/0201163.

\item\label{stef} H. Stefani, D. Kramer, M. Maccallum, C. Hoenselaers,
E. Herlt, {\it Exact Solutions of Einstein's Field Equations} (2nd ed.),
Cambridge University Press, Cambridge (UK), 2003.

\item\label{dub} M. Dubois-Violette, M. Henneaux, Lett. Math. Phys. {\bf
49} (1999) 245;  Commun.\ Math.\ Phys.\  {\bf 226} (2002) 393
[arXiv:math.qa/0110088].

\item\label{10r} A. Taub, Ann. Math. {\bf 53} (1951)
472; \\ E. Newman, L. Tamburino, T. Unit, J. Math. Phys. {\bf 4}
(1963) 915.

\item\label{11r} C. Misner, J. Math. Phys. {\bf 4} (1963) 924.

\item\label{13r} F.~W.~Hehl, P.~Von Der Heyde, G.~D.~Kerlick and
J.~M.~Nester, Rev.\ Mod.\ Phys.\  {\bf 48} (1976) 393; \\
R.~T.~Hammond, Rept.\ Prog.\ Phys.\  {\bf 65} (2002) 599.

\item\label{14r} M.~Tsamparlis, Phys.\ Rev.\ D {\bf 24} (1981) 1451.

\item\label{15r} S.~Capozziello, G.~Lambiase and C.~Stornaiolo, Annalen
Phys.\  {\bf 10} (2001) 713 [arXiv:gr-qc/0101038];\\
S.~Capozziello, V.~F.~Cardone, E.~Piedipalumbo, M.~Sereno and
A.~Troisi, Int.\ J.\ Mod.\ Phys.\ D {\bf 12} (2003) 381
[arXiv:astro-ph/0209610].

\item\label{16r} D.~Lynden-Bell and M.~Nouri-Zonoz, Rev.\ Mod.\ Phys.\ 
{\bf 70} (1998) 427\\ {}[arXiv:gr-qc/9612049]; \\
M.~Nouri-Zonoz and D.~Lynden-Bell, ``Gravomagnetic Lensing by NUT
Space,''\\ {}arXiv:gr-qc/9812094;\\
J.~Q.~Shen, ``Dynamics of gravitomagnetic charge,''
arXiv:gr-qc/0301100.

\een

\end{document}